\documentclass[prd,aps,preprint,showpacs,nofootinbib,superscriptaddress,tightenlines]{revtex4-1}
\usepackage{amsmath}
\usepackage{epsfig}
\usepackage{url}
\newcommand{\ben}{\begin{eqnarray}}
\newcommand{\een}{\end{eqnarray}}
\newcommand{\nnu}{\nonumber\\}
\newcommand{\bef}{\begin{figure}[htb]\centering}
\newcommand{\eef}{\end{figure}}

\begin{document}

\title{Single Spin Asymmetry Scaling in the Forward Rapidity Region at RHIC}

\author{Zhong-Bo Kang}
\email{zkang@bnl.gov}
\affiliation{RIKEN BNL Research Center, 
                   Brookhaven National Laboratory, 
                   Upton, NY 11973, USA}

\author{Feng Yuan}
\email{fyuan@lbl.gov}
\affiliation{Nuclear Science Division, 
                   Lawrence Berkeley National Laboratory, 
                   Berkeley, CA 94720, USA} 
\affiliation{RIKEN BNL Research Center, 
                   Brookhaven National Laboratory, 
                   Upton, NY 11973, USA}
                   
\begin{abstract}
We investigate the scaling properties in inclusive hadron production and the associated
single transverse spin asymmetry in the forward rapidity region at RHIC. We find that
the spin-averaged experimental data in both $pp$ and $dAu$ collisions 
demonstrates a transverse-momentum-dependent geometric scaling. 
We introduce the transverse momentum dependent 
Collins fragmentation function to study the scaling of the single transverse spin 
asymmetries. The general feature of the scaling analysis is consistent with 
the experimental observations, in particular, for the transverse
momentum dependence of the spin asymmetries at RHIC energy.
We further propose to probe the saturation scale of nucleus by measuring the
spin asymmetry normalized by that in $pp$ scattering at low transverse momentum.
\end{abstract}


\maketitle

\section{Introduction}

Since its operation, the forward rapidity region in $pp$ and $dAu$
collisions at the Relativistic Heavy Ion Collider (RHIC) has
provided important opportunities to study novel hadronic physics
phenomena, from the single transverse spin asymmetry (SSA) in $pp$
collisions \cite{Adams:2003fx,:2008qb,:2008mi,Lee:2007zzh} 
to the small-$x$ gluon saturation in $dAu$ collisions \cite{Arsene:2004ux,Adams:2006uz,Braidot:2010zh,Adare:2011sc}. Both physics has attracted great attentions in the last few years.

The nuclear suppression observed in hadron production in the
forward rapidity region in $dAu$ collisions at RHIC has been interpreted as
the consequence of the saturation physics at small-$x$ \cite{Kharzeev:2003wz,Albacete:2003iq,Boer:2007ug,Baier:2005dz,Albacete:2010bs,Albacete:2010pg}. An
effective theory of the color-glass-condensate (CGC) has been applied to
describe these phenomena \cite{Iancu:2003xm}. Alternative approaches have also been
proposed \cite{Qiu:2004da,Guzey:2004zp,Kopeliovich:2005ym,Frankfurt:2007rn}. Meanwhile the large single transverse spin asymmetries
in $pp$ collisions found at RHIC have been studied from various
approaches \cite{Kouvaris:2006zy,Anselmino:1994tv}, while the underlying mechanism 
remains unclear \cite{Kang:2011hk}. 
In this paper, we will extend the CGC formalism to include the spin
effects, in particular, in the fragmentation process. Early
attempts have been reported in Ref.~\cite{Boer:2006rj}, where they focused on the spin
dependent quark distribution contribution.

One important aspect in the context of the color-glass-condensate
is the so-called geometric scaling \cite{GolecBiernat:1998js,Stasto:2000er}. This is better understood as that 
the unintegrated gluon distribution $N_F(x,q_\perp)$ can be written as a single function
of $q_\perp^2/Q_s^2(x)$ with $Q_s(x)$ the saturation scale. 
As a result, the structure function of deep inelastic
scattering can be expressed as a single function of $\tau=Q^2/Q_s^2$ 
with $Q$ the virtuality of the exchanged photon \cite{Stasto:2000er}.
It is remarkable that the experimental data from HERA support
this geometric scaling. Since then, this idea has been applied to many
hadronic processes involving small-$x$ parton distributions, including
some experimental data relevant to the forward region at RHIC.

In the forward rapidity region of $pp$ (or $pA$) collisions at
high energy, the incoming parton from the projectile scatters on
the target. Because of high density of the partons from the target
at small $x$ (or in large nucleus), this parton goes through
multiple scattering before it fragments into the final state
hadron. The CGC formalism takes into account these multiple
interaction effects in a systematic manner. In particular, the
quantum evolution at small-$x$ in this formalism has been well
established and plays a very important role to understand the
suppression phenomena observed at RHIC \cite{Dumitru:2005gt}. 

We further introduce the transverse momentum dependence in the
fragmentation functions to understand the single transverse spin
asymmetries by introducing the spin-dependent fragmentation
function - the Collins function \cite{Collins:1992kk}. Previous model
studies have claimed that this effect is too small to make any
significant contribution to the SSA in forward hadron production
in $pp$ collisions \cite{Anselmino:2004ky}. However, we will find that the CGC formalism
provides a simple factorized form for the spin-dependent cross
section, from which we find that the Collins mechanism indeed
plays an important role. Of course, the limitation of the current
knowledge of the quark transversity distribution and the Collins
fragmentation function do not allow us to make reliable
predictions. However, from the positivity bounds of both
functions, we find that it is possible to explain the large single
spin asymmetries in hadron production observed by the experiments
at RHIC.

The rest of this paper is organized as follows. In Sec.~II, we analyze 
the inclusive hadron production in the forward rapidity region in $pp$ and
$pA$ collisions following the CGC approach. 
We will show that the experimental data from RHIC
demonstrates a transverse momentum dependent geometric scaling for
both $pp$ and $dAu$ collisions. 
In Sec.~III, we introduce
the transverse momentum dependence in the fragmentation process to
understand the single spin asymmetries in $pp$ collisions. In
particular, the transverse momentum dependence of $A_N$ indicates
the importance of details in the hadronization. We conclude our
paper in Sec.~IV.

\section{Transverse Momentum Dependent Geometric Scaling}

In the forward rapidity region, the hadron production in $pp$ and
$pA$ collisions at high energy is dominated by the valence quark
and gluon from the projectile. The energetic parton penetrates in
the nucleon (nucleus) target with multiple scattering and then
fragments into the final state hadron. 
We follow the CGC approach to study the differential cross section of
hadron production $p+p({\rm or~} A)\to h+X$
in the forward rapidity region \cite{Albacete:2010bs, Dumitru:2005gt},
\ben
\frac{d\sigma}{dy_h d^2P_{h\perp}}=\frac{K}{(2\pi)^2} \int_{x_F}^1
\frac{dz}{z^2}x_1 q(x_1)N_F(x_2, k_\perp=P_{h\perp}/z) D_{h/q}(z) \ ,
\label{unpol}
\een
where $K$ is a possible $K$-factor to absorb the higher order corrections, 
$x_F=P_{h\perp}/\sqrt{s}\, {\rm exp}(y_h)$, and $x_1=x_F/z$ is the momentum fraction of the projectile carried by
the incoming quark, $z$ the momentum fraction of the quark carried
by the final state hadron, $x_2=x_1\,{\rm exp}(-2y_h)$ is the momentum fraction of the
target participating in the hard scattering. We have also
suppressed the scale dependence in various factors in the above
formula, for which the precise forms will depend on the
next-to-leading order perturbative corrections.

$N_F(x_2, k_\perp)$ is the unintegrated  gluon distribution, and is given by the two-dimensional Fourier transform of the imaginary part of the forward dipole-target scattering amplitude
in the fundamental representation:
\ben
N_F(x_2, k_\perp) = \int d^2 r e^{-i k_\perp\cdot r}
\left[1-{\mathcal N}_F(r, Y=\ln(x_0/x))\right],
\een
with $r$ the dipole size. The characteristic feature of ${\mathcal N}_F(r, Y)$ is its ``geometric scaling'', i.e., ${\mathcal N}_F(r, Y)$ only depends on the dimensionless quantity $r^2 Q_s^2(x)$ \cite{GolecBiernat:1998js,Stasto:2000er}. Thus its Fourier transform has ``geometric scaling'': $k_\perp^2 N_F(x_2, k_\perp)$ depends only on the dimensionless ratio $k_\perp^2/Q_s^2(x)$. This property will manifest itself in the differential cross section of hadron production as we will show below.

For the large-$x$ quark and gluon distribution, an important power counting  can be applied to study the general behavior for the
differential cross section of the above process \cite{Gunion:1973nm,Brodsky:1994kg}. In particular,
for the region where the quark distribution dominates, we have
$q(x)\sim (1-x)^3$. Furthermore, fragmentation function also has
power counting rule, $D_{h/q}(z)\sim (1-z)$, where the power behavior comes
from the current parameterization of the fragmentation function
for the charged and neutral pions \cite{FFs} \footnote{The original power
counting would predict $(1-z)^2$ behavior for the spin-averag
fragmentation functions~\cite{Brodsky:1994kg}. This would
lead to a power behavior of $(1-x_F)^6$ instead of $(1-x_F)^5$ in Eq.~(\ref{formula}). The 
current data in Fig.~1 can be described by both parameterizations. Future experimental
data at very forward region shall be able to distinguish them.}. Combining the
above power counting analysis, we obtain the following power
behavior for the differential cross section,
\ben
P_{h\perp}^2\frac{d\sigma}{dy_h d^2P_{h\perp}}=(1-x_F)^5{\cal
F}\left(\frac{P_{h\perp}^2}{Q_s^2(x_2)}\right)\ ,
\label{formula}
\een
where the saturation scale $Q_s$ depends on the target and the
momentum fraction $x_2$ carried by the gluon. The geometric scaling 
(as a function of $P_{h\perp}^2/Q_s^2$ only) comes from the geometric
scaling of the unintegrated gluon distribution for 
$k_\perp^2 N_F(x,k_\perp)$ as mentioned above.  

Similar geometric scaling for inclusive hadron production
at central rapidity has been observed in \cite{McLerran:2010ex, Tribedy:2010ab}, 
where it is the  differential cross section alone exhibits the geometric scaling. 
On the contrary,  we have to multiply a factor of $P_{h\perp}^2$ to the differential 
cross section to show the geometric scaling. The geometric scaling
derived in \cite{McLerran:2010ex, Tribedy:2010ab} relies on the 
$k_T$-factorization (implicitly), and both incoming hadrons contribute to the saturation effects in
the particle production at mid-rapidity, where the saturation scales 
from the two incoming hadrons are in the same order at mid-rapidity. 
However, in our analysis, we have a dilute projectile 
scattering on a dense target, where the saturation only comes from 
the dense target and the geometric scaling is manifest in the first
place. Of course, as a result, the differential cross section has to be
multiplied by a factor of $P_{h\perp}^2$ to demonstrate the geometric 
scaling.
\bef 
\vskip 0.2in
\psfig{file=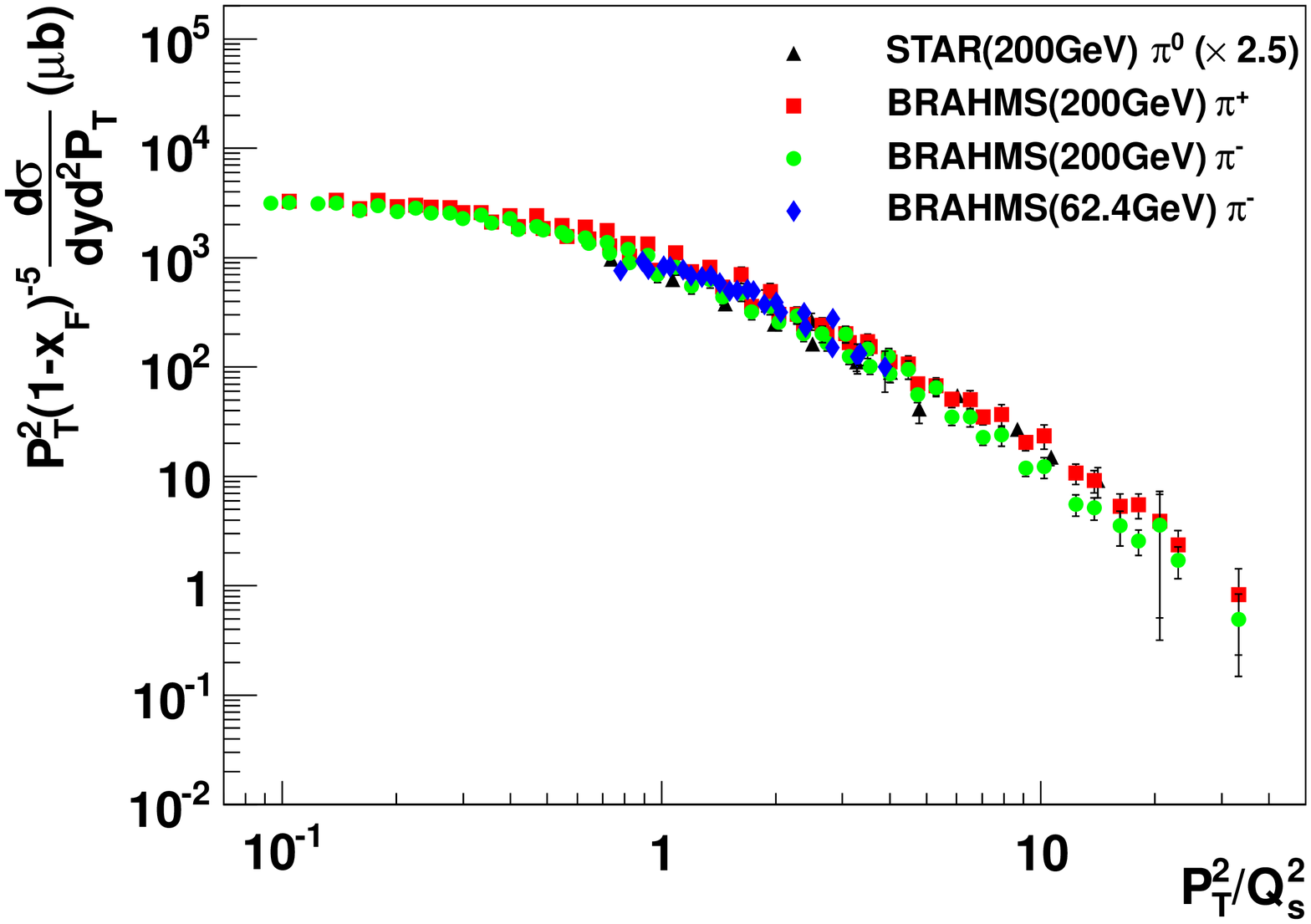,width=3.1in} 
\hskip 0.2in
\psfig{file=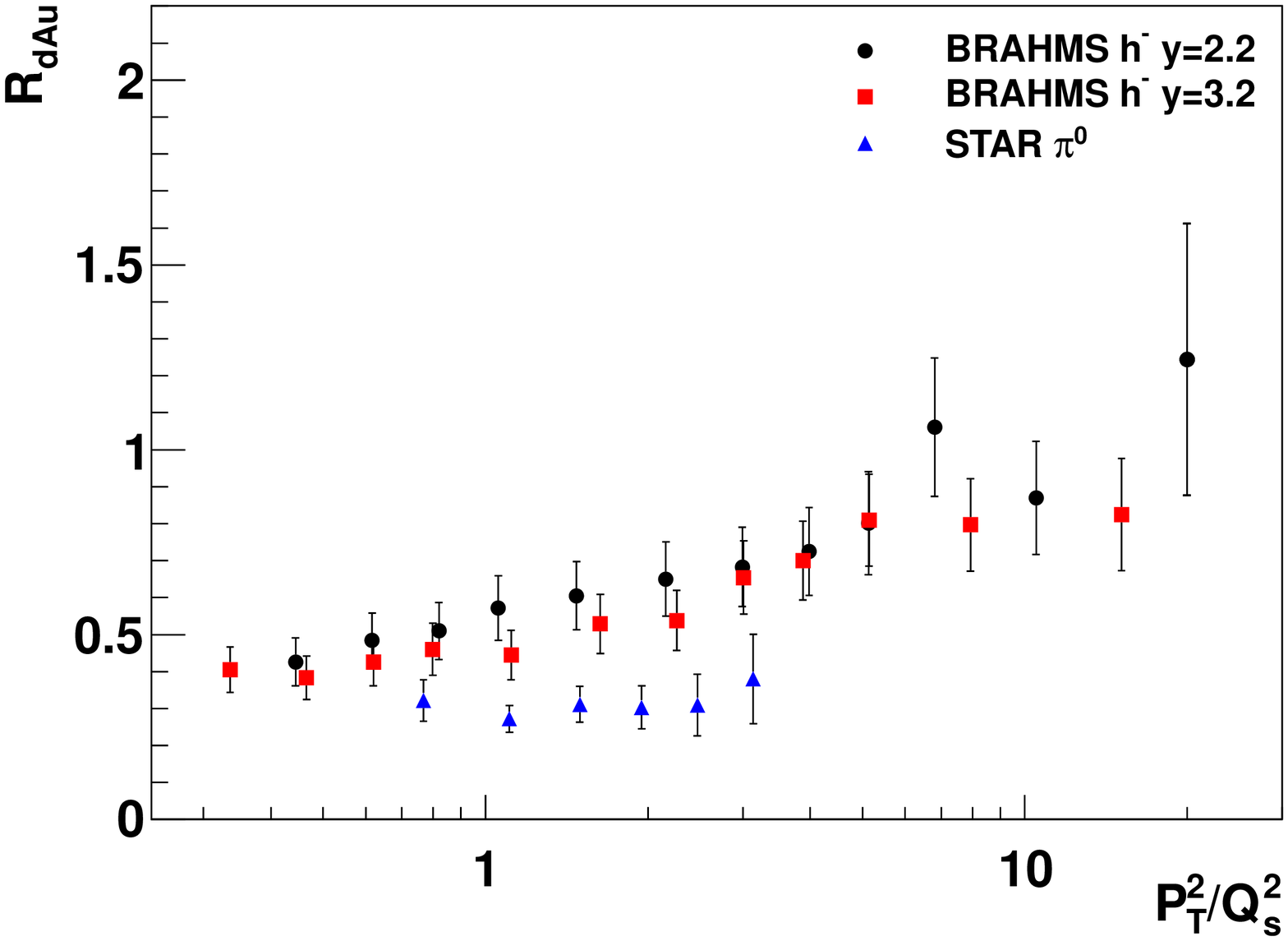,width=3.0in} 
\caption{Geometric scaling of
the differential cross sections (left) and the nuclear modification factor $R_{\rm dAu}$
(right) plotted as a function of $P_{h\perp}^2/Q_s^2$. Data are from
BRAHMS \cite{Arsene:2004ux} and STAR \cite{Adams:2006uz} Collaborations at RHIC.} 
\label{scaling}
\eef

In Fig.~\ref{scaling}, we plot the RHIC data in $pp$ collisions the
differential cross section,
\ben
(1-x_F)^{-5}P_{h\perp}^2
\frac{d\sigma(pp\to h+X)}{dy_h d^2P_{h\perp}}
\een
as a function of $P_{h\perp}^2/Q_s^2(x_2)$, where $x_2$ is
calculated as $x_2=P_{h\perp}^2/(x_1 s)\approx P_{h\perp}^2/(x_F S)$, and
the saturation scale $Q_s^2(x_2)=Q_0^2
(x_0/x_2)^{\lambda}$ with $Q_0=1$ GeV, $x_0=3\cdot 10^{-4}$, and $\lambda=0.28$ \cite{GolecBiernat:1998js}.
In this plot, we have included the data for $\pi^\pm$ from BRAHMS
collaboration \cite{Arsene:2004ux} and $\pi^0$ from the STAR collaboration \cite{Adams:2006uz}
at $\sqrt{s}=200$ GeV. $\pi^\pm$ data are from two rapidity bins $y_h=2.95$ and $y_h=3.3$, whereas those for
$\pi^0$ cover $y_h=3.3, 3.8, 4.0$. We have also included data for $\pi^-$
production at lower energy $\sqrt{s}=62.4$ GeV from BRAHMS at $y_h=2.7$ 
and $y_h=3.3$ \cite{Arsene:2004ux}.
For $\pi^0$ production, we have multiplied a factor $2.5$, which is consistent with
the overall $K$-factor $K\sim 0.4$ found in \cite{Albacete:2010bs}.
From this plot, we see clearly that the hadron production in the forward 
rapidity region demonstrates the geometric
scaling. 

It is also interesting to check the geometric scaling when we
compare the $pA$ and $pp$ collisions in the forward rapidity
region~\cite{Armesto:2004ud}. 
Here, the nuclear modification factor $R_{\rm pA}$ is measured,
\ben
R_{\rm pA}=\frac{\left.dN^{pA\to h+X}\right/dy_hd^2P_{h\perp}}{N_{coll}
\left.dN^{pp\to h+X}\right/dy_hd^2P_{h\perp}} \ ,
\een
where $N_{coll}$ represents the number of binary collisions in the
selected nucleon-nucleus scattering. Many factors cancel out in
the above ratio. Therefore, according to Eq.~(\ref{formula}) 
we will have the following geometric scaling behavior,
\ben
R_{\rm pA}={\cal R}_b\left(\frac{P_{h\perp}^2}{Q_s^2(x_2)}\right)\ ,
\een
where again the saturation scale $Q_s$ depends on the momentum
fraction of the nucleus target involved in the scattering process
and is determined by the kinematics of $P_{h\perp}$ and rapidity
$y_h$. In the above equation the subscript $b$ indicates that the
ratio also depends on the centrality of the collision, which
determines the change of the saturation scale from $pp$ collision
to $pA$ collision at a given impact parameter. Of course, the
magnitude ${\cal R}$ depends on the overall normalization as well,
including the parameter $N_{coll}$.

As an illustration, we plot the $R_{\rm dAu}$ for hadron
production at $\sqrt{s}=200$ GeV in the forward rapidity region,
where we include the BRAHMS data for the negative charged hadron
$h^-$ at two rapidities: $y_h=2.2$ and $y_h=3.3$, and the STAR data
for neutral pion at $y_h=4.0$. Clearly, the two rapidity sets from
BRAHMS demonstrate the geometric scaling behavior, whereas the
STAR data are a little off the scaling curve of the negative charged
hadrons. This indicates additional mechanism for very forward
hadron suppression in $dA$ collisions~\cite{Frankfurt:2007rn}.
It will be very interesting to further test this
geometric scaling in the future experiments at RHIC and the LHC.

The power behavior for the differential cross section has been
observed in various scattering energies, and has been used as
a practice to understand the forward hadron production. 
However, the geometric scaling only appears when we hit
the saturation region at small-$x$ of the target. For example, 
we notice that the forward hadron production at low energy
fixed-target experiments does not show the same scaling behavior \cite{LloydOwen:1980cp}. 
This tells us that geometric scaling is not yet reached in these energies,
and the production mechanism differs. Similar conclusions have been drawn
from the comparison between the RHIC data and those from the previous 
fixed-target experiments in the collinear factorization approach \cite{Bourrely:2003bw,deFlorian:2005yj}. 

\section{Single Spin Asymmetry in the Forward Region}

In this section, we extend the above formalism to
calculate the single spin asymmetry in the forward region. 
For this purpose, we need to introduce the transverse
momentum dependence either in the incoming parton distribution
from the projectile (polarized proton)~\cite{Boer:2006rj} and/or the fragmentation
function. Here we concentrate on the contribution
from the fragmentation function.

In the forward rapidity, introducing the transverse momentum dependence in
the fragmentation function in Eq.~(\ref{unpol}), we have
the spin-averaged differential cross section
\ben
\frac{d\sigma}{dy_h d^2P_{h\perp}}=\frac{K}{(2\pi)^2} \int_{x_F}^1
\frac{dz}{z^2}\int d^2P_{hT} x_1q(x_1)N_F(x_2, k_\perp) D_{h/q}(z, P_{hT}),
\label{unpolpt}
\een
where $P_{hT}$ is the transverse momentum of the final state hadron relative 
to the fragmenting quark jet, thus the momentum of the final state hadron in Lab frame 
can be written as: ${\mathbf P}_{h\perp}\approx z\, {\mathbf k}_\perp+{\mathbf P}_{hT}$. On the other hand, the spin-dependent differential cross section 
for $p^\uparrow + p({\rm or~} A)\to h+X$ can
be written as \cite{Yuan:2008tv}
\ben
\frac{d\Delta\sigma}{dy_h d^2P_{h\perp}}=\frac{K}{(2\pi)^2}
\int_{x_F}^1 \frac{dz}{z^2}\int d^2P_{hT}I(S_\perp, P_{hT})
x_1h(x_1) N_F(x_2, k_\perp)\delta\hat{q}(z,
P_{hT}), 
\label{main} 
\een 
where $h(x_1)$ is the quark transversity, $\delta\hat{q}(z, P_{hT})$ is the Collins function that is
related to the Trento convention \cite{Bacchetta:2004jz} as: 
$\delta\hat{q}(z, P_{hT})=-H_1^\perp/{zM_h}$ with $M_h$ the final-state hadron mass. The azimuthal angle dependence is encoded in $I(S_\perp, P_{hT})$ and is given by 
\ben
I(S_\perp, P_{hT})=\epsilon^{\alpha\beta}S_\perp^\alpha\left[P_{hT}^\beta
-\frac{n\cdot P_{hT}}{n\cdot P_J}P_J^\beta\right], 
\een 
where $\epsilon^{\alpha\beta}$ is the 2-dimensional Levi-Civita tensor with $\epsilon^{12}=1$, 
$n$ is a unit vector along the momentum of the unpolarized hadron, and
$P_J$ is the momentum of the fragmenting quark jet. In the forward
rapidity region, $I(S_\perp, P_{hT})$ can be reduced to the
following form, 
\ben 
I(S_\perp, P_{hT})=|S_\perp||P_{hT}|\sin(\phi_h-\phi_s), 
\label{angdep} 
\een 
where $\phi_s$ is the azimuthal angle for the spin vector $S_\perp$ in the Lab frame, 
$\phi_h$ is the azimuthal angle of the produced hadron momentum $P_{h\perp}$ relative to 
the fragmenting quark jet.

The single transverse spin asymmetry is defined as 
\ben
A_N=\left.\frac{d\Delta\sigma}{dy_hd^2P_{h\perp}}\right/\frac{d\sigma}{dy_h d^2P_{h\perp}}.
\een
To be consistent with the experimental definition we choose the coordinate 
system where hadron transverse momentum $P_{h\perp}$ along $x$-direction, the spin
vector $S_\perp$ along $y$-direction, and the polarized beam along
$z$-direction. 

Before we present the numeric estimates, we would like to study the generic
feature for the single spin asymmetry in the above approach. 
We will take a simple model for the unintegrated gluon distribution and the
Collins fragmentation function to  study the transverse momentum 
behavior of the single spin asymmetry. First, we will examine
the single spin asymmetry at relative low transverse momentum at order
of the saturation scale: $P_{h\perp}^2\sim Q_s^2$. 
We know that the unintegrated gluon distribution
has saturation behavior at low transverse momentum. As a convenient 
model, we follow the GBW parameterization \cite{GolecBiernat:1998js}
\ben 
N_F(x,q_\perp)\sim \frac{1}{Q_s^2}e^{-{q_\perp^2}/{Q_s^2}}, 
\een
which has geometric scaling.
For the fragmentation function we choose a simple Gaussian distribution,
\ben
D_{h/q}(z, p_\perp)\sim \frac{1}{\Delta^2}e^{-{p_\perp^2}/{\Delta^2}},
\een 
where $\Delta^2$ represents the width of the transverse momentum dependence in the
fragmentation function. By performing the
transverse momentum integration in Eq.~(\ref{unpolpt}), we will find that the spin-averaged cross 
section has the following transverse momentum behavior, 
\ben
\frac{d\sigma}{dy_hd^2P_{h\perp}}\sim \frac{1}{Q_s^2+\Delta^2}e^{-{P_{h\perp}^2}/{(Q_s^2+\Delta^2)} },
\label{avgpt}
\een
where we limit the above results at small transverse momentum region 
of order of $Q_s$. If the saturation scale $Q_s$ is much larger than the transverse
momentum width $\Delta$ in the fragmentation function, we will find that the transverse
momentum dependence in the fragmentation function does not change
the geometric scaling discussed in the last section.

Now we turn to the spin-dependent differential cross section. For the Collins function, we assume it  also has a Gaussian form \cite{Yuan:2008tv},
\ben
\delta\hat{q}(z, p_\perp^2)\sim \frac{1}{(\Delta^2-\delta^2)^{3/2}}e^{-{p_\perp^2}/{(\Delta^2-\delta^2)}},
\een
with a slight difference in the Gaussian width $\Delta^2-\delta^2$ to satisfy the positive 
bound. Substituting the above equation to 
the differential cross section formula in Eq.~(\ref{main}),
one obtains,
\ben
\frac{d\Delta\sigma}{dy_hd^2P_{h\perp}}\propto \frac{P_{h\perp}\sqrt{\Delta^2-\delta^2}}{(Q_s^2+\Delta^2-\delta^2)^2}e^{-\frac{P_{h\perp}^2}{Q_s^2+\Delta^2-\delta^2}}\ ,
\een
From the above results, we find that the single spin asymmetry behaves as,
\ben
A_N(P_{h\perp})&\propto &\frac{P_{h\perp}(Q_s^2+\Delta^2)\sqrt{\Delta^2-\delta^2}}{(Q_s^2+\Delta^2-\delta^2)^2}
e^{-\frac{P_{h\perp}^2}{Q_s^2+\Delta^2-\delta^2}
+\frac{P_{h\perp}^2}{Q_s^2+\Delta^2}} 
\nnu
&\approx & \frac{P_{h\perp}\Delta}{Q_s^2+\Delta^2}e^{-\frac{\delta^2P_{h\perp}^2}{(Q_s^2+\Delta^2)^2}}
\nnu
&\approx &\frac{P_{h\perp}\Delta}{Q_s^2}e^{-\frac{\delta^2P_{h\perp}^2}{(Q_s^2)^2}}  \ ,
\label{lowpt}
\een
where we have made reasonable assumptions: $Q_s^2\gg \Delta^2\gg \delta^2$. 
The above result indicates that the asymmetry vanishes when $P_{h\perp}\to 0$,
and it also depends on the transverse momentum width in the fragmentation 
function. Certainly, if there is no transverse momentum dependence, the whole
effects will vanish. Furthermore, the spin asymmetry also decreases with the
saturation scale. This is because the fragmentation effect is suppressed 
if we increase the saturation scale (see also Eq.~(\ref{avgpt})).
From the above simple analysis, we find that the spin asymmetry in general will
have broader distribution as a function of $P_{h\perp}$. 

Moreover, it is interesting to note that the double ratio of the spin asymmetries
between $p^{\uparrow}A$ and $p^{\uparrow}p$ scatterings scales as
\ben
\left.\frac{A_N^{pA\to h}}{A_N^{pp\to h}}\right|_{P_{h\perp}^2 \ll Q_s^2}
\approx 
\frac{Q_{sp}^2}{Q_{sA}^2}e^{\frac{P_{h\perp}^2\delta^2}{Q_{sp}^4}}\ ,
\een
at small transverse momentum, where we have assumed that the saturation 
scale for nucleus is much 
larger than that for the nucleon at the same kinematics. This is the most 
interesting result from the scaling analysis. The ratio of the spin asymmetry 
is inversely proportional to the saturation scale in the limit
of $P_{h\perp}\to 0$.  This can be used as an important probe for the
saturation scale of the gluon distribution in the nuclei target.

Similarly, we can estimate the large transverse momentum behavior for the spin
asymmetries, where the unintegrated gluon distribution behaves as 
\ben
N_F(x,q_\perp)\sim \frac{Q_s^2}{q_\perp^4}.
\een
If we still assume that the fragmentation function can be parametrized as a Gaussian 
form, we will find out,
\ben
A_N(P_{h\perp})\propto \frac{2P_{h\perp}\sqrt{\Delta^2-\delta^2}}{P_{h\perp}^2+4\Delta^2} \ ,
\label{highpt}
\een
where the factor $4$ comes from the power of the unintegrated gluon distribution at large transverse momentum. 
The asymmetry decreases as $1/P_{h\perp}$ at large transverse momentum as 
expected. However, the rate of the decreasing is strongly affected by the relative size between
$P_{h\perp}^2$ and $\Delta^2$. This additional $4\Delta^2$ modification is slightly different from the usual $1/P_{h\perp}$ power counting results \cite{Kouvaris:2006zy}. It comes from the effects of the transverse momentum dependent fragmentation function in the spin-averaged cross section which was neglected previously \cite{Kouvaris:2006zy}.

Furthermore, we notice that the saturation scale dependence cancels out between
the spin-averaged and the spin-dependent cross sections, and the asymmetry 
does not depend on the saturation scale. 
As a consequence, the double ratio will approach
\ben
\left.\frac{A_N^{pA\to h}}{A_N^{pp\to h}}\right|_{P_{h\perp}^2 \gg Q_s^2}\approx 1\ .
\een
The above results are very interesting observations, and it will be important to test
these predictions in the future experiments \cite{decadal}.
\bef
\vskip 0.1in
\psfig{file=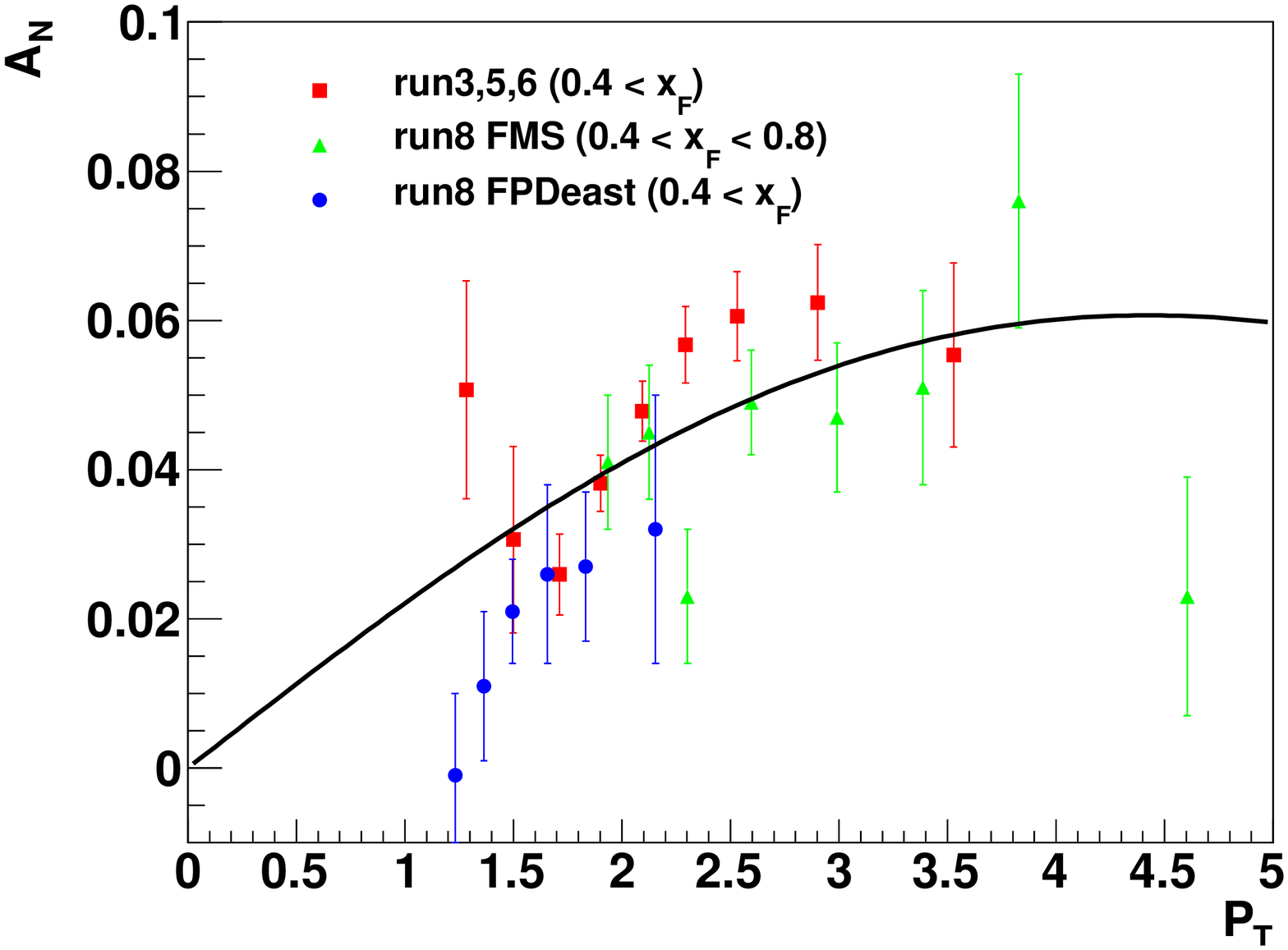, width=3.0in}
\hskip 0.2in
\psfig{file=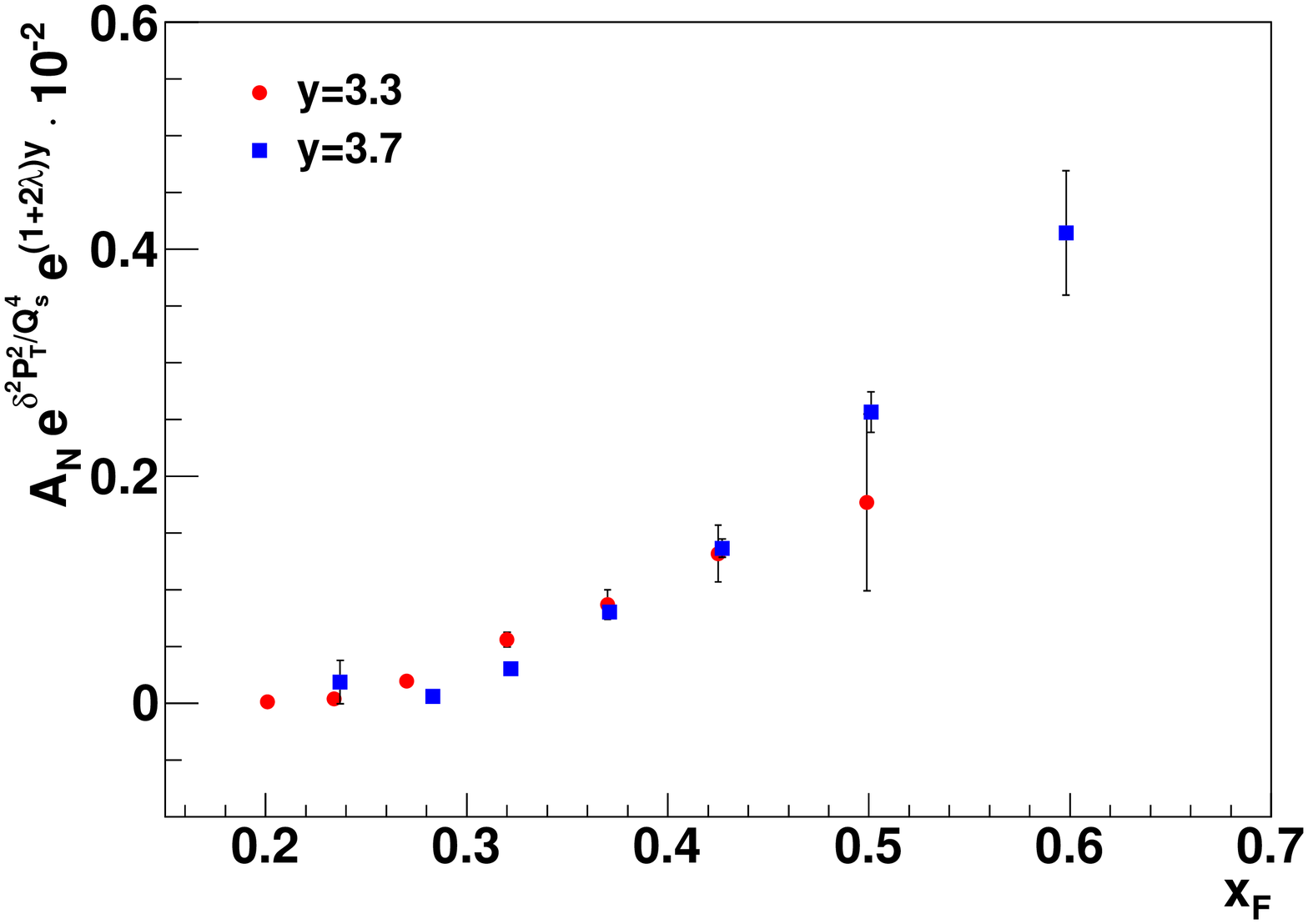, width=3.15in}
\caption{Left: $P_{h\perp}$-dependence of the single spin asymmetry. We use Eq.~(\ref{lowpt}) to fit the STAR experimental data \cite{:2008qb, Drachenberg:2009nd}. Right: We plot $A_N e^{\delta^2P_{h\perp}^2/Q_s^4} e^{(1+2\lambda)y}$ as a function of $x_F$ - the generalized $x_F$-scaling for the single spin asymmetry according to Eq.~(\ref{xfscaling}). We have chosen $\delta=0.16$ GeV to demonstrate the scaling. The data is from STAR \cite{:2008qb}.}
\label{ptdep}
\eef

Comparing Eqs.~(\ref{lowpt}) and (\ref{highpt}), we find that the maximum of the single spin asymmetry will be
at $P_{h\perp}^2\sim Q_s^4/\delta^2$. Since $\delta^2$ is a small number, we would find that the spin asymmetry drops at a relatively large transverse momentum,
which seems to be consistent with the observations at RHIC experiments \cite{:2008qb,phenix-pt}. In Fig.~\ref{ptdep} (left), we show 
the comparison of Eq.(\ref{lowpt}) with the experimental data from the STAR collaboration \cite{:2008qb, Drachenberg:2009nd}, where we have chosen $Q_s^2\approx 1.0$ GeV$^2$, which is roughly the average saturation scale in this experimental kinematic region. $\delta\approx 0.16$ GeV is adjusted to fit the experimental data. Although the normalization of the asymmetry depends on 
$\Delta^2$ and the size of the Collins effect, including the quark transversity distribution
and the Collins fragmentation function, the transverse momentum dependence from our general 
analysis is consistent with the experimental data.

Meanwhile, from Eq.~(\ref{lowpt}) and realizing $P_{h\perp}=x_F \sqrt{S} e^{-y_h}$, we find
\ben
A_N \sim x_F^{(1+\lambda)} e^{-(1+2\lambda)y_h} e^{-\delta^2P_{h\perp}^2/Q_s^4} {\cal F}\left(x_F\right),
\een
where ${\cal F}\left(x_F\right)$ represents any additional $x_F$ dependence for the single spin asymmetry. Therefore, the following quantity
\ben
A_N e^{\delta^2P_{h\perp}^2/Q_s^4} e^{(1+2\lambda)y_h} \sim x_F^{(1+\lambda)} {\cal F}\left(x_F\right),
\label{xfscaling}
\een
will be a universal function of $x_F$, independent of the rapidities. In Fig.~\ref{ptdep} (right), we plot STAR $\pi^0$ single spin asymmetry data for the quantity defined in Eq.~(\ref{xfscaling}) as a function of $x_F$ for two rapidities $y_h=3.3$ and $y_h=3.7$. We find that for relatively small $\delta \lesssim 0.2$ GeV (which is consistent with our approximation), the data seem to be consistent with the $x_F$-scaling. It will be interesting to test this scaling in the future experiments.
Note that we derive this interesting scaling purely from the fragmentation process, but it could also come from the distribution function (Sivers function) in the polarized nucleon \cite{Kouvaris:2006zy}. However, the complicated structure for the Sivers function makes such a study nontrivial~\cite{Kang:2011hk}. 

\bef 
\vskip 0.1in
\psfig{file=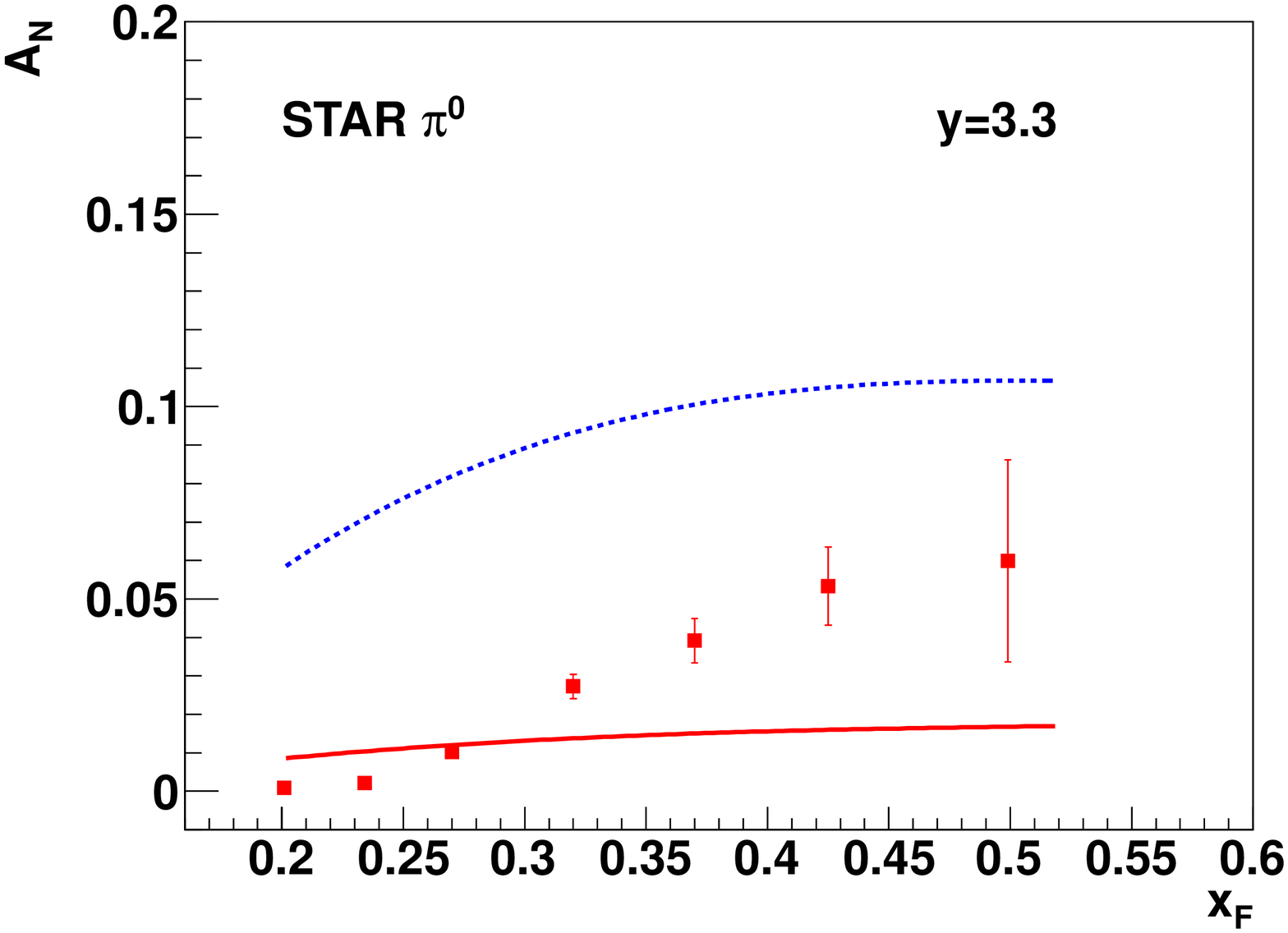, width=3.1in} 
\hskip 0.2in
\psfig{file=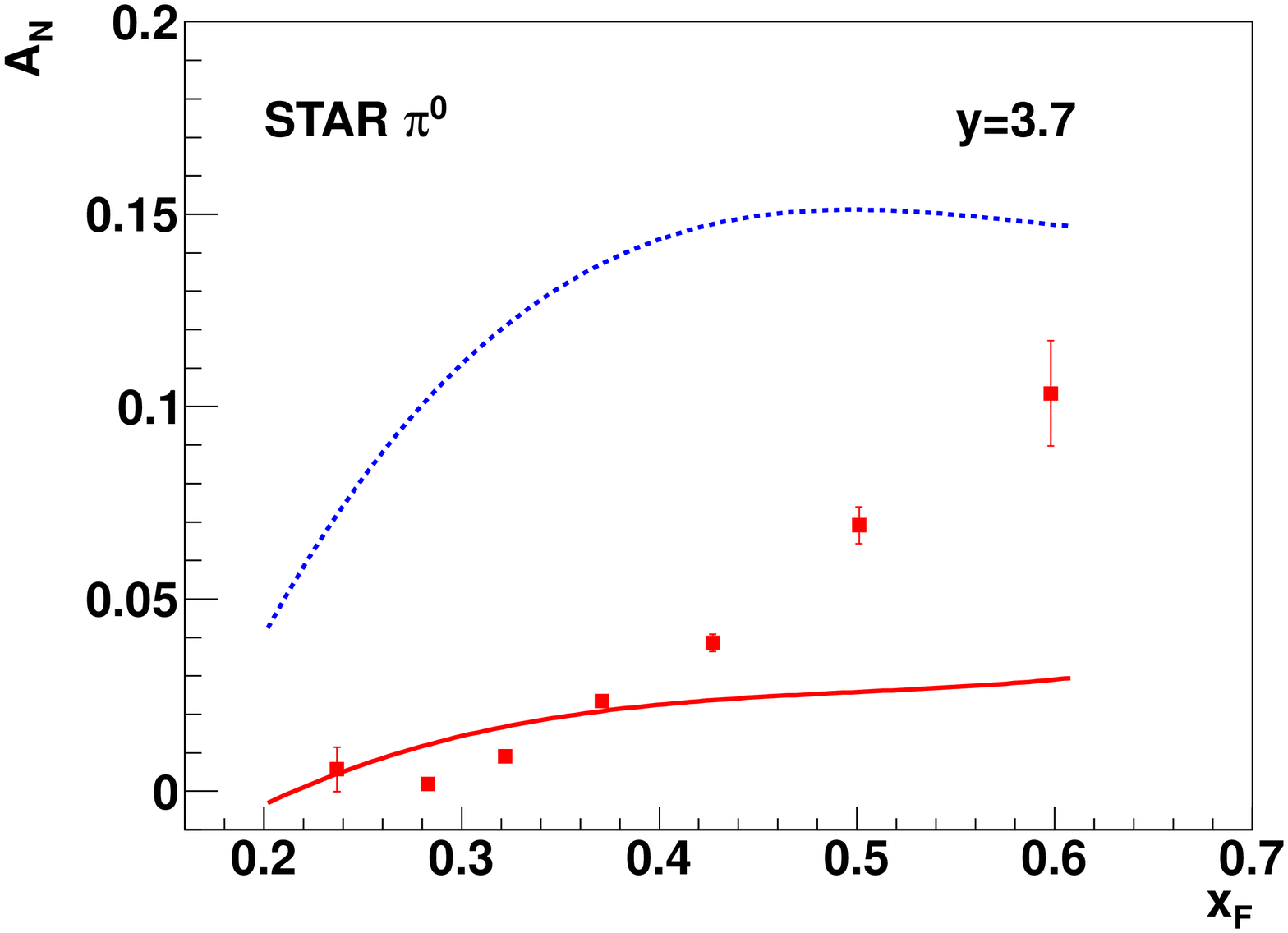, width=3.1in} 
\caption{The comparison with the STAR experimental data at $\sqrt{s}=200$ GeV
for rapidity $y_h=3.3$ (left) and $y_h=3.7$ (right). 
The solid lines are using the transversity from \cite{Martin:1997rz,Vogelsang:2005cs} and the Collins function from \cite{Yuan:2008tv, Vogelsang:2005cs}. 
The dashed curves are the upper bound as explained in the text.} 
\label{starxf} 
\eef
\bef
\vskip 0.1in
\psfig{file=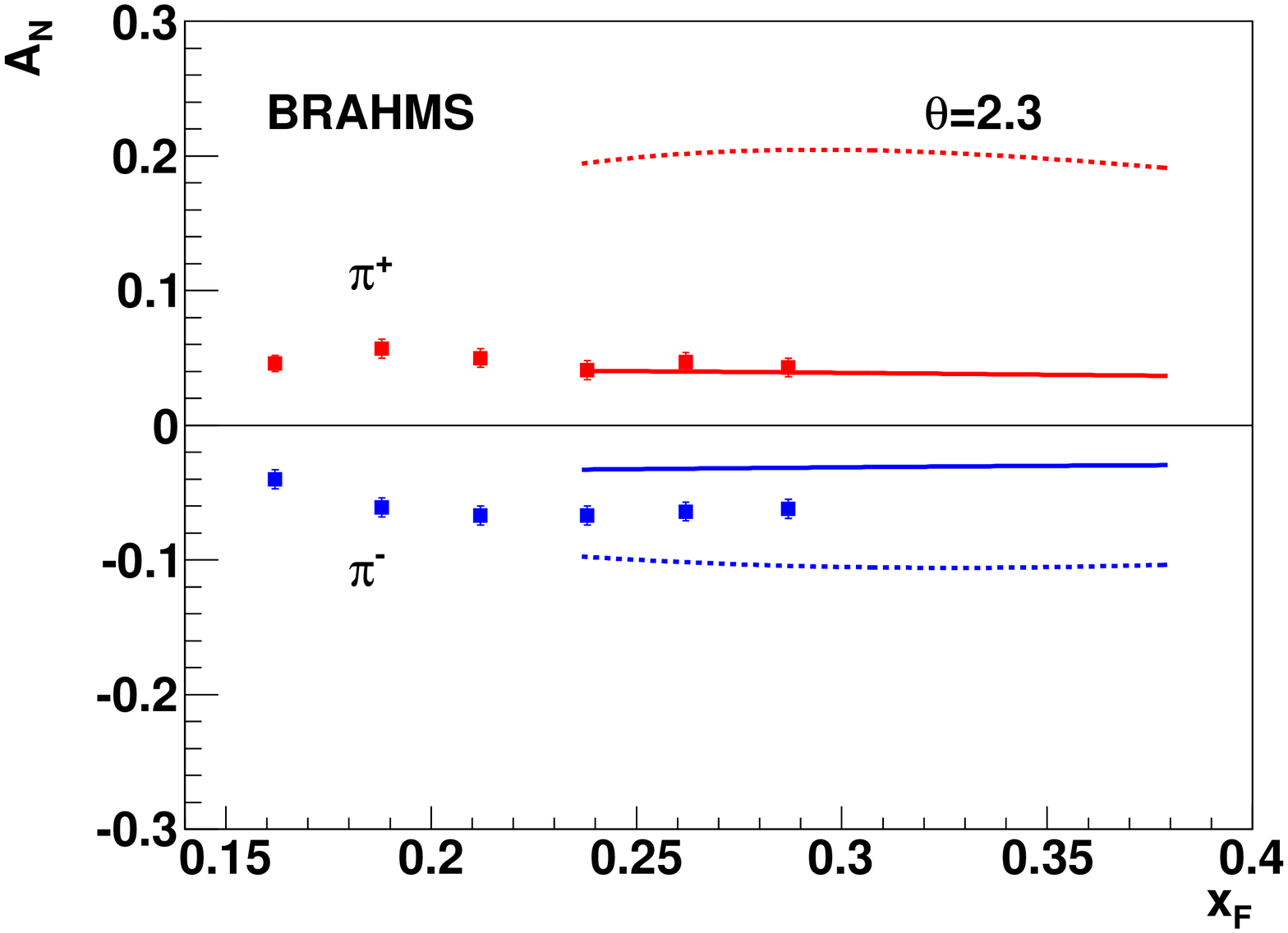, width=3.1in}
\hskip 0.2in
\psfig{file=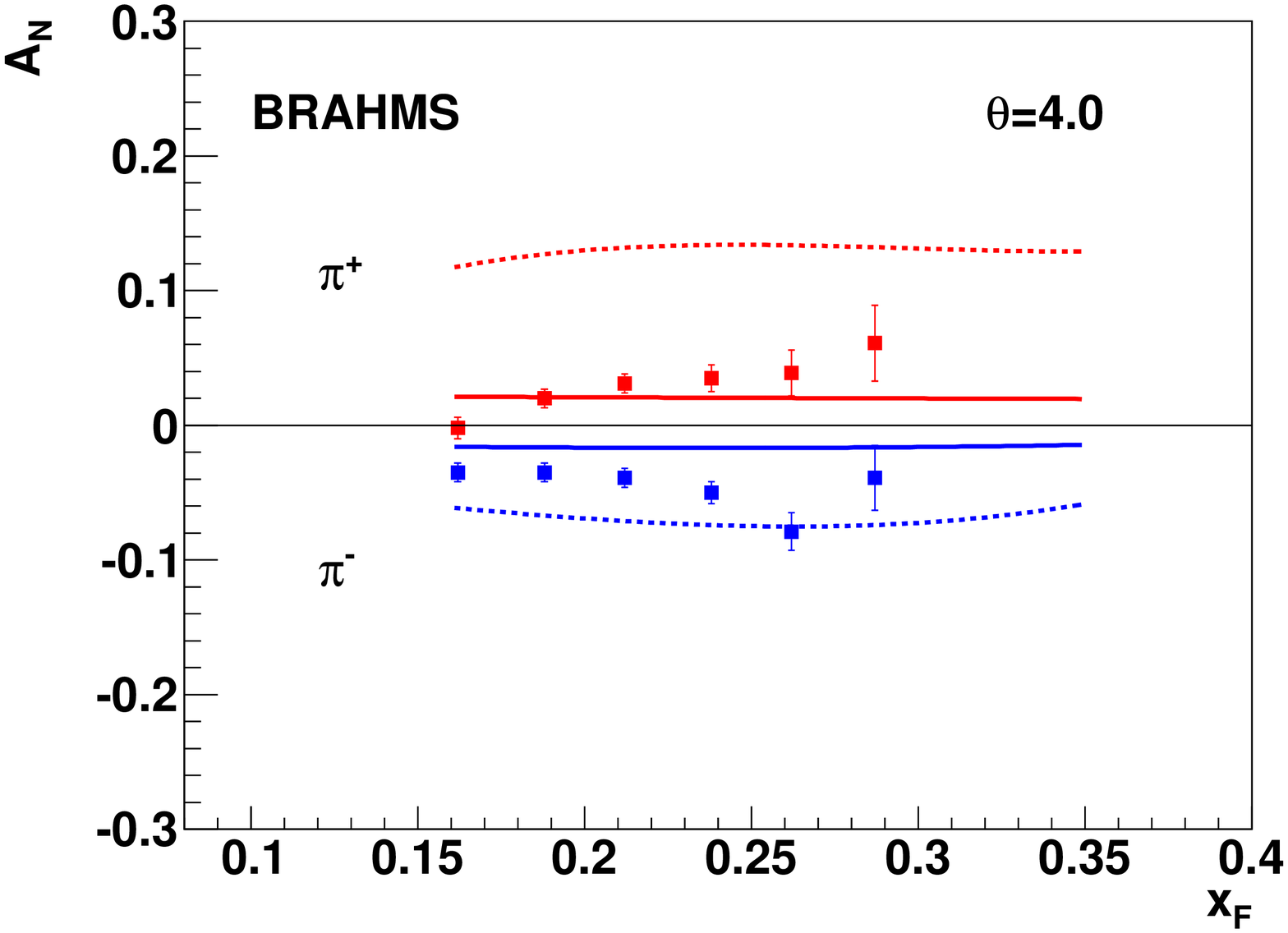, width=3.1in} 
\caption{The comparison with BRAHMS experimental data at $\sqrt{s}=200$ GeV. The curves
are the same as explained in Fig.~\ref{starxf}.}
\label{brahmsxf}
\eef
Let us now come back to make some numerical estimates for the single spin asymmetry. For this purpose, we will use the unintegrated gluon distribution in Ref.~\cite{Albacete:2010bs}, which comes from solving Balitsky-Kovchegov equation with running-coupling corrections. We will
choose the quark transversity in \cite{Martin:1997rz,Vogelsang:2005cs}, and adopt the transverse momentum dependent unpolarized fragmentation function and Collins function in \cite{Yuan:2008tv, Vogelsang:2005cs}, 
\ben
D_{h/q}(z, P_{hT})&=&\frac{1}{\pi\langle p_\perp^2\rangle} e^{-P_{hT}^2/\langle p_\perp^2\rangle} D_{h/q}(z),
\\
\delta\hat{q}(z, P_{hT})&=&\frac{2}{\left(\pi\langle p_\perp^2\rangle\right)^{3/2}}e^{-P_{hT}^2/\langle p_\perp^2\rangle} \delta\hat{q}^{(1/2)} (z),
\een
where $\langle p_\perp^2\rangle=0.2$ GeV$^2$, and $\delta\hat{q}^{(1/2)} (z)$ is the so-called half-moment of the Collins function given in \cite{Yuan:2008tv, Vogelsang:2005cs}. The numerical estimates for the single spin asymmetry of $\pi^0$ are shown in Fig.~\ref{starxf} as solid lines, and also compared with the experimental data from STAR collaboration \cite{:2008qb}. Similar comparison for the charged pions are shown in Fig.~\ref{brahmsxf} with the experimental data from BRAHMS collaboration \cite{Lee:2007zzh}. 

From these plots, it seems that the predictions are below the experimental data. However, it is important to keep in mind that there are uncertainties in both the quark transversity distribution and the Collins fragmentation function. For example, the quark transversity is constrained using experimental data from HERMES collaboration, only up to $x\lesssim 0.3$ \cite{Airapetian:2004tw}, thus we are lacking of information on the transversity in the interested region at RHIC forward direction $x > 0.3$. The situation for the Collins function is slightly better since we have complimentary information from both semi-inclusive deep inelastic scattering at HERMES \cite{Airapetian:2004tw} and $e^+e^-$ collisions at BELLE \cite{Abe:2005zx}. But one still needs to understand how to incorporate the evolution of the Collins function, which will affect the Gaussian approximation and the Gaussian width used above~\cite{Idilbi:2004vb}. Realizing the limitation of the current knowledge for both quark transversity and Collins function, we will set up a bound for the single spin asymmetry by using the positivity bounds for both functions.

For transversity, we saturate the Soffer bound \cite{Soffer:1994ww}, which is derived from positivity \cite{Artru:2008cp}. Following Ref.~\cite{Anselmino:2004ky}, we choose
\ben 
h_u(x)=\frac{1}{2}\left[u(x)+\Delta
u(x)\right], 
\qquad 
h_d(x)=-\frac{1}{2}\left[d(x)+\Delta
d(x)\right],
\een
where $\Delta q(x)$ is the helicity distribution function, $h_q(x)$ is the quark transversity. For the Collins function, we have
\ben 
P_{hT}\delta\hat{u}^{\pi^+}(z, P_{hT})&=&- D_u^{\pi^+}(z, P_{hT}),
\\
P_{hT}\delta\hat{d}^{\pi^+}(z, P_{hT})&=&D_d^{\pi^+}(z, P_{hT}).
\een
where the signs are chosen such that we have the maximum single spin asymmetry for the charged pions \cite{Anselmino:2004ky}. For $\pi^0$, we rely on the isospin symmetry. In Figs.~\ref{starxf} and \ref{brahmsxf}, the dashed lines are the upper bound for the single spin asymmetries. From these plots, we see the Collins mechanism in CGC formalism is large enough to explain the observed asymmetries. If the Collins mechanism is the dominant contribution to the single spin asymmetries
\cite{Kang:2011hk}, one should be able to extract the quark transversity and Collins function by using the RHIC data. 

\section{Summary}
We study the scaling properties in the inclusive single hadron production and the associated single transverse spin asymmetries in the forward rapidity region at RHIC.
For the spin-averaged differential cross sections in both $pp$ and $pA$ collisions, we find a transverse momentum dependent geometric scaling. These scalings are clearly seen in the spin-averaged experimental data in both $pp$ and $dAu$ collisions. We further introduce the transverse momentum dependence in the fragmentation function to investigate the scaling of the single transverse spin asymmetry associated with the Collins mechanism.  The general behavior of the transverse momentum dependence of the single spin asymmetry is consistent with the trend observed in the experiments at RHIC. We also find that the double ratio of the spin asymmetries in $pA$ and $pp$ collisions is inversely proportional to the saturation scale in the limit of $P_{h\perp}\to 0$. This can be used to probe the saturation scale in the nucleus by measuring the spin asymmetry normalized by that in $pp$ scattering at low transverse momentum.

With the limited knowledge for the relevant functions - quark transversity and Collins function, we set up an upper bound for the single spin asymmetry. We find that the bound is certainly large enough to explain the experimental data at RHIC. 

\section*{Acknowledgments}
We thank L.~McLerran and R.~Venugopalan for helpful discussions, and thank J.~L.~Albacete for providing us their unintegrated gluon distribution used in our numerical estimate. We also thank L.~Bland, L.~Eun, S.~Heppelmann, J.~H.~Lee, A.~Ogawa, and F.~Videbaek for the discussions
on the experimental data.
This work was supported in part by the U.S. Department of Energy
under Grant No.~DE-AC02-05CH11231.
We are grateful to RIKEN, Brookhaven National Laboratory,
and the U.S. Department of Energy (Contract No.~DE-AC02-98CH10886)
for supporting this work.


\end{document}